\documentclass[twocolumn]{aastex62}

\usepackage{amsmath}
\usepackage{appendix}

\graphicspath{{./}{figures/}}
\shortauthors{Callister et al.}

\begin{document}

\title{A First Search for Prompt Radio Emission from a Gravitational-Wave Event}

\correspondingauthor{T. Callister}
\email{thomas.a.callister@gmail.com}

\author{Thomas A. Callister}
\affiliation{LIGO Laboratory, California Institute of Technology, 1200 E. California Blvd, Pasasdena, CA 91125 USA}

\author{Marin M. Anderson}
\affiliation{California Institute of Technology, 1200 E. California Blvd, Pasasdena, CA 91125 USA}

\author{Gregg Hallinan}
\affiliation{California Institute of Technology, 1200 E. California Blvd, Pasasdena, CA 91125 USA}

\author{Larry R. D'addario}
\affiliation{California Institute of Technology, 1200 E. California Blvd, Pasasdena, CA 91125 USA}

\author{Jayce Dowell}
\affiliation{Department of Physics and Astronomy, University of New Mexico, Albuquerque, NM 87131 USA}

\author{Namir E. Kassim}
\affiliation{Naval Research Laboratory, Washington, DC 20375, USA}

\author{T. Joseph W. Lazio}
\affiliation{Jet Propulsion Laboratory, California Institute of Technology, 4800 Oak Grove Dr, Pasadena, CA 91109, USA}

\author{Danny C.\ Price}
\affiliation{Department of Astronomy, 501 Campbell Hall \#3411, University of California Berkeley, Berkeley CA 94720}
\affiliation{Centre for Astrophysics \& Supercomputing, Swinburne University of Technology, Hawthorn, VIC 3122, Australia}

\author{Frank K. Schinzel}
\affiliation{Department of Physics and Astronomy, University of New Mexico, Albuquerque, NM 87131 USA}
\affiliation{National Radio Astronomy Observatory, P.O. Box O,
Socorro, NM 87801}

\begin{abstract}
Multimessenger observations of the binary neutron star merger GW170817 have enabled the discovery of a diverse array of electromagnetic counterparts to compact binary mergers, including an unambiguous kilonova, a short gamma-ray burst, and a late-time radio jet.
Beyond these counterparts, compact binary mergers are additionally predicted to be accompanied by prompt low-frequency radio emission.
The successful observation of a prompt radio counterpart would be immensely valuable, but is made difficult by the short delay between the gravitational-wave and prompt electromagnetic signals as well as the poor localization of gravitational-wave sources.
Here, we present the first search for prompt radio emission accompanying a gravitational-wave event, targeting the binary black hole merger GW170104 detected by the Advanced LIGO and Virgo gravitational-wave observatories during their second (O2) observing run.
Using the Owens Valley Radio Observatory Long Wavelength Array (OVRO-LWA), we search a $\sim900\,\mathrm{deg}^2$ region for transient radio emission within approximately one hour of GW170104, obtaining an upper limit of $2.5\times10^{41}\,\mathrm{erg}\,\mathrm{s}^{-1}$ on its equivalent isotropic luminosity between 27-84 MHz.
We additionally discuss plans to target binary neutron star mergers in Advanced LIGO and Virgo's upcoming O3 observing run.
\vspace{1cm}
\end{abstract}

%%%%%%%%%%%%%%%%%%%%%%
\section{Introduction}
\label{sec:intro}
%%%%%%%%%%%%%%%%%%%%%%

The detection of the binary neutron star merger GW170817 has heralded the era of multi-messenger astronomy \citep{BNS_gw,BNS_mma,BNS_gbm,BNS_optical,BNS_radio,BNS_xray}.
Observed in both gravitational waves and virtually every electromagnetic band, this event yielded an extraordinary amount of information, including measurements of nucleosynthesis in kilonovae \citep{BNS_kilonova1,BNS_kilonova2}, new insights into gamma-ray burst mechanisms \citep{BNS_jet1,BNS_jet2}, constraints on the neutron star equation of state \citep{BNS_eos1,BNS_eos2}, and even an independent measurement of the Hubble constant \citep{BNS_H0,BNS_H02}.

Beyond the electromagnetic counterparts associated with GW170817, binary neutron star mergers are predicted to be accompanied by prompt radio emission \citep{Usov2000,Hansen2001,Pshirkov2010,Lai2012,Lyutikov2013,Totani2013,Ravi2014,Metzger2016,Wang2016,Lyutikov2018,Wang2018}.
Unlike the late-time radio afterglow associated with GW170817, due to the interaction of relativistic ejecta with the ambient medium, the theorized prompt radio emission is generated by processes internal to the merging objects themselves.
In particular, prompt emission may take the form of a short (likely sub-second) coherent radio pulse generated near the instant of merger.

The detection of prompt radio emission from a binary neutron star would yield an immense amount of information, probing the binary's immediate magnetic environment near the time of merger, tracing properties of the intergalactic medium, and offering rapid $\sim$arcminute constraints on the progenitor's location.
However, the observation of prompt emission is made difficult by several factors \citep{ChuProspects,YanceyProspects,KaplanProspects}.
First, gravitational-wave detectors provide only poor localization of gravitational-wave sources.
Even for the three-detector Advanced LIGO-Virgo network, the median binary neutron star localization is expected to be $120-180\,\mathrm{deg}^2$ during the upcoming O3 observing run \citep{Prospects}.
Second, low-frequency prompt emission released at time of merger may arrive at Earth as little as one minute after the gravitational-wave signal, slowed only by free electrons encountered during propagation.
Searches for prompt radio emission are therefore typically limited by the latency with which gravitational-wave candidates are announced -- notices released more than minutes after a gravitational wave's arrival may well come too late.

All previous searches for prompt radio emission have targeted short gamma-ray bursts \citep{Anderson2018,Kaplan2015,Obenberger2014,Bannister2012} or were carried out too late to detect any prompt emission that may have been present \citep{LWA-170817,MWA-170817}.
Here, we perform the first search for prompt radio emission coincident with a gravitational-wave signal, using the Owens Valley Radio Observatory Long Wavelength Array (OVRO-LWA).
Observing between 27 and 84 MHz, the OVRO-LWA consists of 288 dual-polarization antennas spanning $\sim1.5$ km.
Cross correlation of 256 antennas with the Large-Aperture Experiment to Detect the Dark Age (LEDA) correlator provides all-sky imaging with 24 kHz frequency resolution and $\sim$10 arcmin spatial resolution \citep{leda,Anderson2018,Eastwood2018}.

The OVRO-LWA is uniquely suited to the challenge of detecting prompt radio emission.
Its nearly hemispherical field of view can capture much of the LIGO-Virgo localization region within a single image.
Additionally, the OVRO-LWA is equipped with a 24-hour buffer to which visibilities can be continuously written.
This alleviates (although does not eliminate; see Sect. \ref{sec:future}) the need for rapid LIGO-Virgo notices.
Provided that a notice is released within one day of the gravitational-wave event, the relevant on-source data can be retrieved from the buffer and written to disk.

Although the OVRO-LWA was observing at the time of GW170817, the binary neutron star merger occurred below the OVRO-LWA's horizon \citep{BNS_gw}.
We therefore cannot place any observational limits on the prompt radio emission associated with GW170817.
Instead, we report the results of a search for prompt radio emission associated with the binary black hole merger GW170104 \citep{gw170104,gw170104_supplement}.

Although stellar-mass binary black hole mergers are generally not expected to yield electromagnetic transients, the tentative \textit{Fermi-GBM} detection of gamma-rays associated with the binary black hole merger GW150914 \citep{GW150914,fermi_2016} has sparked new interest in possible counterparts to LIGO/Virgo's binary black hole events.
In particular, binary black holes might conceivably generate electromagnetic transients if one or more of the black holes is charged \citep{Liebling2016,Liu2016,Zhang2016,Fraschetti2018}, in the presence of a circumstellar or circumbinary disk \citep{Perna2016,deMink2017}, or in the case of black hole ``twins'' born from the collapse of a single massive star \citep{Loeb2016}.
Although the statistical significance of the \textit{Fermi-GBM} candidate remains under debate \citep{LyutikovFermi2016,Greiner2016,Savchenko2016,Connaughton2018}, the plethora of models predicting electromagnetic counterparts makes binary black hole mergers an interesting (if speculative) observational target.

While valuable in its own right, the search for prompt radio emission from GW170104 additionally serves as a powerful proof-of-principle.
GW170104 exemplifies the challenges facing detection of prompt radio emission.
First, its accompanying localization is poor, spanning a significant fraction of the sky.
Second, the LIGO/Virgo alert announcing the detection of GW170104 was released hours after the gravitational wave event, long after the expected arrival of any prompt radio emission.
Despite these challenges, we place stringent upper limits on the prompt radio luminosity of GW170104, demonstrating the capability of the OVRO-LWA to follow up future compact binary mergers.

In Sect. \ref{sec:gw170104} below, we begin by describing the gravitational-wave signal GW170104 as well as the OVRO-LWA data collection and initial reduction.
In Sect. \ref{sec:dedispersion}, we then describe the search for a dispersed radio transient, and in Sect. \ref{sec:luminosity} present upper limits on the radio flux and luminosity of GW170104.
Finally, in Sect. \ref{sec:future} we discuss future prospects for the follow-up of gravitational-wave candidates.

%%%%%%%%%%%%%%%%%%%%
\section{GW170104 and OVRO-LWA Observations}
\label{sec:gw170104}
%%%%%%%%%%%%%%%%%%%%

% 17/01/04 16:49:56 GMT == 1167583814 (GPS) == Jan 04 2017 16:49:56 UTC 

\begin{figure}
\centering
\includegraphics[width=0.45\textwidth]{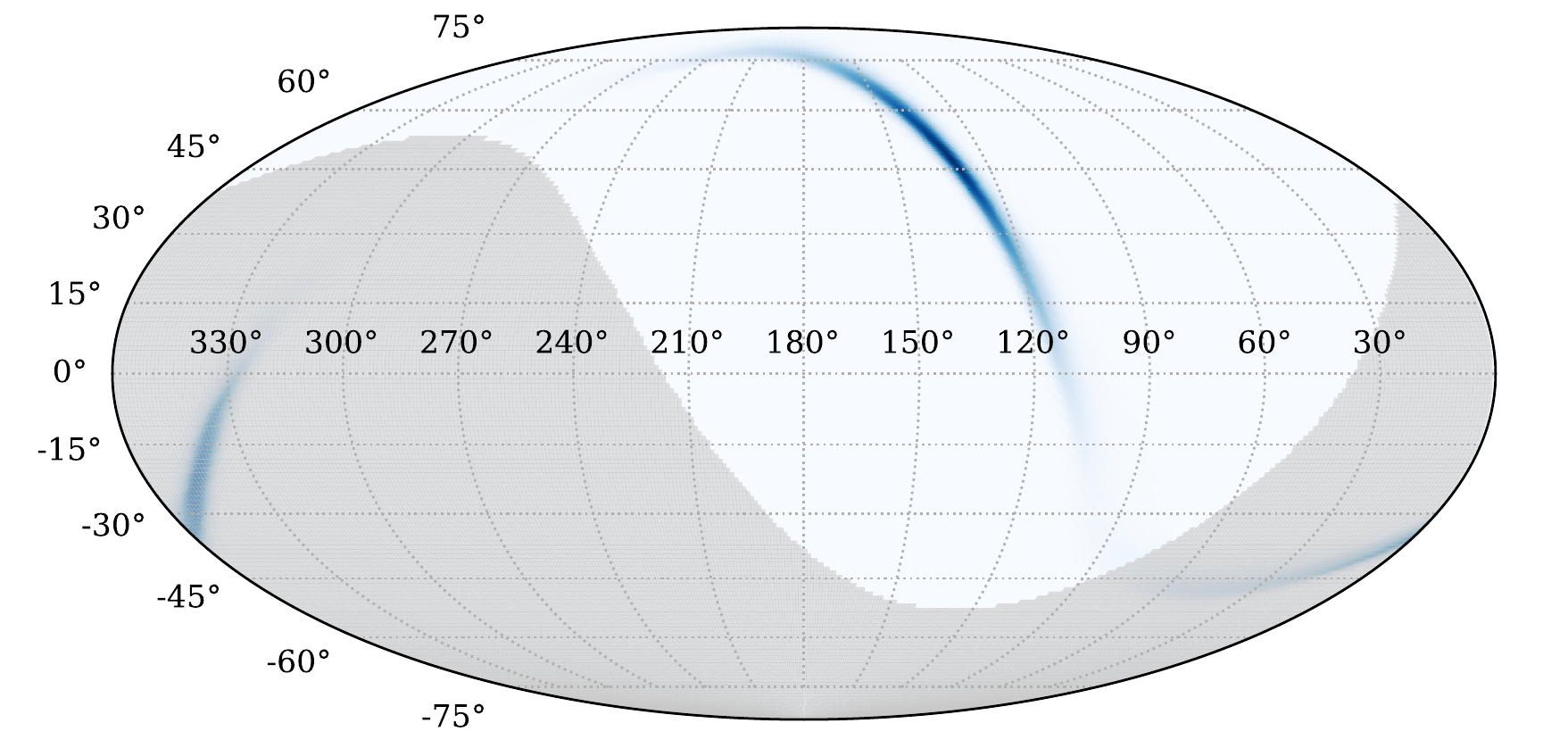}
\caption{
Posterior probability distribution (in blue) on the sky position of the binary black hole merger GW170104.
Also shown is the OVRO-LWA's field of view at GW170104's time of arrival; areas below the OVRO-LWA horizon are shaded in grey.
}
\label{fig:skymap}
\end{figure}

The gravitational-wave signal GW170104 was measured on January 4, 2017 at 10:11:58.6 UTC by the Advanced LIGO experiment \citep{gw170104,gw170104_supplement}.
Arising from a $31+19\,M_\odot$ binary black hole merger, the signal was initially localized to a $\sim1600$ square degree band on the sky (see Fig. \ref{fig:skymap}) and its redshift estimated to be $z=0.18^{+0.08}_{-0.07}$.
Following a delay related to the calibration of Advanced LIGO's Hanford detector, an alert with preliminary event localization was released at 16:49:56 UTC, six hours after the gravitational wave's arrival \citep{LigoLowLatency}.

At this time, the OVRO-LWA was under continuous operation, temporarily storing 13\,s integrations in a continuously-overwritten 24-hour buffer.
Upon receiving the gravitational-wave event notice, buffered data spanning 09:00:03 to 14:11:11 UTC were copied to disk.

Data are flagged on a per antenna, baseline, and channel basis.
We flag antennas showing anomalous autopower spectra, cutting an average of 54 antennas ($\sim38\%$ of visibilities).
An additional 398 baselines are flagged to mitigate cross-talk between adjacent signal paths and eliminate other spurious excess power.
Finally, loud individual channels are automatically flagged to reduce RFI contamination, removing $\sim12\%$ of the 2398 frequency channels.

Cassiopeia (Cas) A and Cygnus (Cyg) A are the brightest sources in the low-frequency radio sky and therefore make opportune calibration sources.
We calibrate our visibility data using a single integration recorded roughly ten hours earlier, at 22:44:04 Jan 03 UTC (21:49:47 local sidereal time), when both Cas A and Cyg A are close to zenith.
The per-channel complex gains of each antenna are determined using a simplified sky model comprising three point sources -- Cas A, Cyg A, and the Sun \citep{Baars1977,Perley2017}.

Following this initial calibration, there persist residual errors due to unmodeled directional variations in antenna gains.
To combat sidelobe contamination arising from these errors, we ``peel'' bright sources, performing an additional direction-dependent calibration and subtraction of these sources \citep{ttcal}.
At the time of GW170104, Cas A and Taurus (Tau) A are the brightest sources in the OVRO-LWA field of view (Cyg A had since set below the horizon).
We peel both Cas A and Tau A, as well as a generic near-field source to remove a stationary noise pattern likely caused by cross-talk between electronics \citep{Eastwood2018}.
Because Cas A is nearly on the OVRO-LWA's horizon, this peeling procedure fails for a small number of integrations; these integrations are manually flagged.

\begin{figure}
\centering
\includegraphics[width=0.45\textwidth]{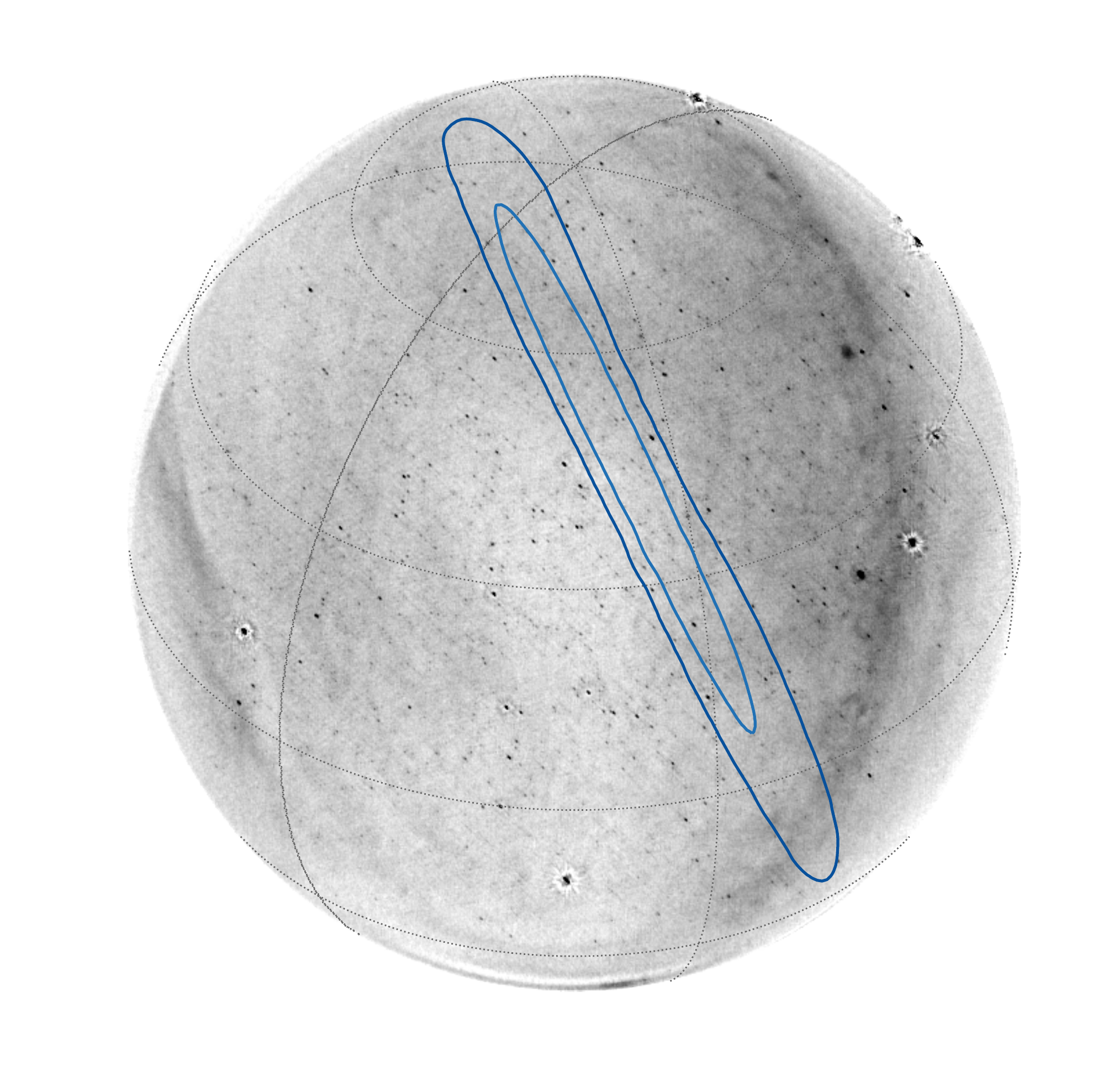}
\caption{
Total intensity image of the OVRO sky from 27 to 84\,MHz in a 13\,s interval centered at 10:11:54.1 UTC, containing GW170104's time of arrival.
The dark and light blue contours show the 95\% and 68\% credible bounds on the location of GW170104's progenitor, respectively, conditioned on the OVRO-LWA's field of view.
}
\label{fig:lwa-sky}
\end{figure}

Figure \ref{fig:lwa-sky} shows a peeled and deconvolved 13\,s image of the OVRO-LWA sky at the time of GW170104 with 0.125\,deg resolution.
Deconvolution is performed using the \texttt{wsclean} algorithm with a Briggs weighting of 0 and a multiscale bias of 0.6 \citep{Offringa2014}.
The blue contours show the 68\% and 95\% credible bounds on the sky location of GW170104's progenitor, restricted to the OVRO-LWA's field of view.
The 95\% credible contour contains 72,556 pixels, each of which we search for a dispersed radio signal.

% 09:00:03 to 14:11:11 UTC 
% Cut to 09:00:03 to 11:12:00 UTC (610 integrations)

%%%%%%%%%%%%%%%%%%%%%%%%%%
\section{Search for a Dispersed Signal}
\label{sec:dedispersion}
%%%%%%%%%%%%%%%%%%%%%%%%%%

\begin{figure*}
\centering
\includegraphics[width=0.95\textwidth]{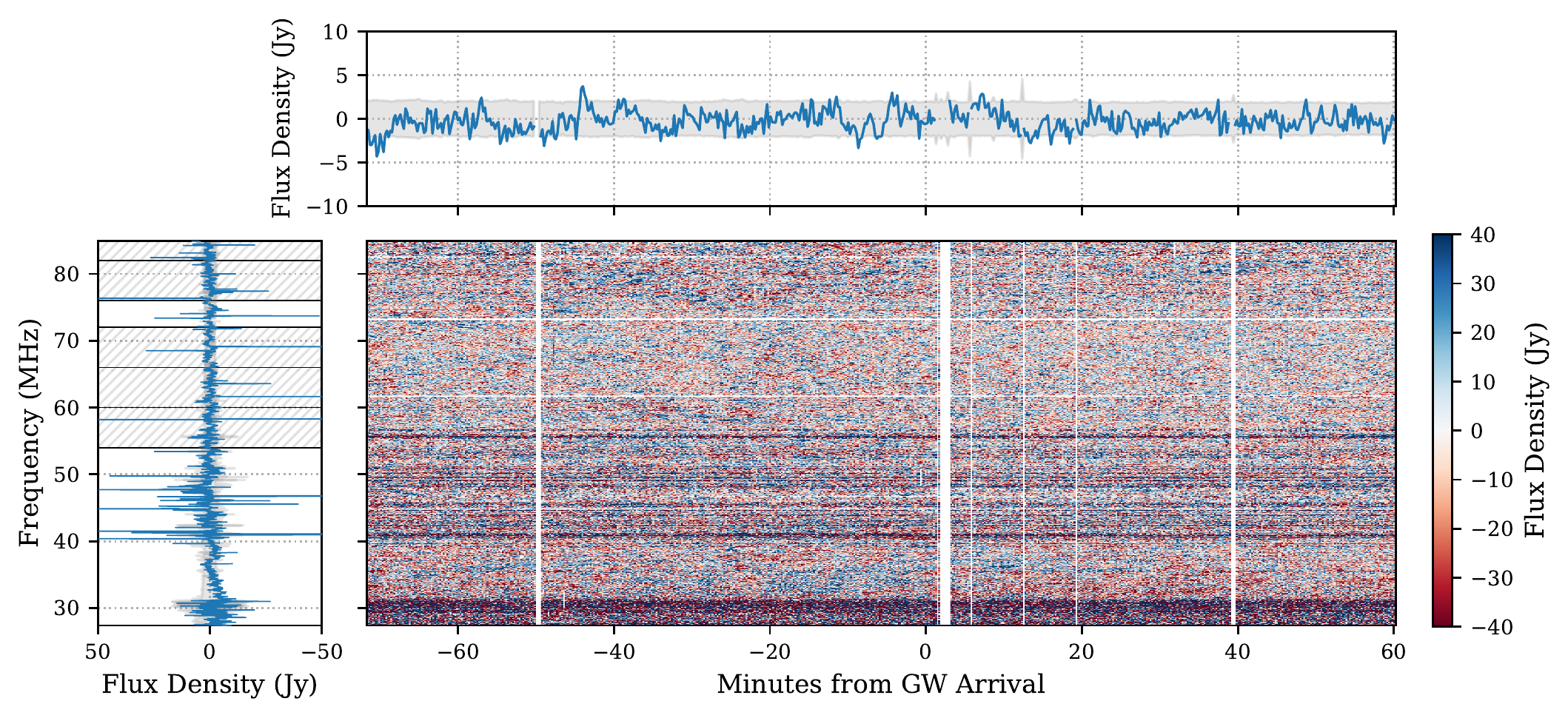}
\caption{
Dynamic spectrum of a randomly-chosen sky location within the GW170104 localization region (Fig. \ref{fig:lwa-sky}), after subtraction of the median flux measured in an annulus surrounding the target location.
White vertical and horizontal bands correspond to times and frequency channels that have been flagged due to excess antenna power or RFI.
The left and upper subplots show the time- and frequency-averaged flux densities, respectively.
The filled grey region within each subplot marks the $\pm 3\sigma$ band as measured in the background annulus.
Broadcast television channels are denoted by hatched regions in the time-averaged spectrum; these channels represent common sources of RFI due to meteor reflection events.
}
\label{fig:spectrogram}
\end{figure*}

\begin{figure*}
\centering
\includegraphics[width=0.8\textwidth]{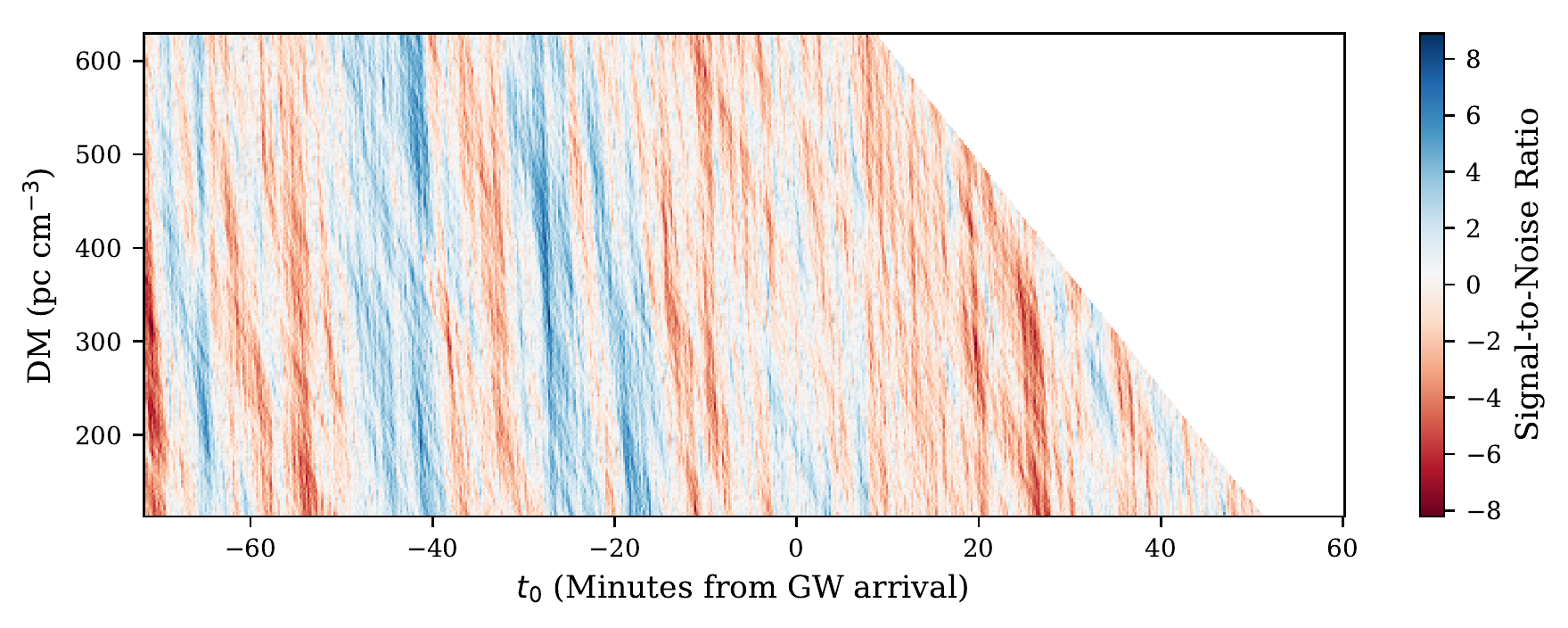}
\caption{
Signal-to-noise ratios (from Fig. \ref{fig:spectrogram}) as a function of dispersion measure DM and the initial time $t_0$ at which a signal is presumed to enter the OVRO-LWA band.
As described in Sect. \ref{sec:dedispersion}, we search for signals with dispersion measures $113\,\mathrm{pc\,cm}^{-3} \leq \mathrm{DM} \leq 630\,\mathrm{pc\,cm}^{-3}$ in a roughly one-hour window around the GW170104's time of arrival.
The blank region on the figure's right-hand side corresponds to time-frequency tracks that extend beyond the duration of our data set.
}
\label{fig:waterfall}
\end{figure*}

Radio waves of frequency $\nu$ propagating through the interstellar and/or intergalactic media experience a dispersion delay
\begin{equation}
\begin{aligned}
t &= \left(\frac{e^2}{2\pi m_e c}\right) \frac{\mathrm{DM}}{\nu^2} \\
    &= \left(4.149\times10^{3}\,\mathrm{s}\right)
            \left(\frac{\mathrm{DM}}{\mathrm{pc\,cm}^{-3}}\right)
            \left(\frac{\nu}{\mathrm{MHz}}\right)^{-2}
\end{aligned}
\label{eq:tDM}
\end{equation}
relative to signals of infinite frequency.
Here, $e$ is the fundamental charge, $m_e$ is the electron mass, $c$ is the speed of light, and the dispersion measure $\mathrm{DM}$ is the integrated column density of free electrons along the wave's path.
The dispersion measure may contain contributions from the immediate environment and/or host galaxy of GW170104, the intergalactic medium, and the interstellar medium of the Milky Way.
Using gravitational-wave constraints on GW170104's redshift and sky location, we bound the dispersion measure of an associated radio transient to $113\,\mathrm{pc\,cm}^{-3} \leq \mathrm{DM} \leq 630\,\mathrm{pc\,cm}^{-3}$ (see Appendix \ref{sec:dm}),
corresponding to time delays ranging from 640--3600\,s at the bottom of the OVRO-LWA band.
We therefore analyze data up to one hour after the gravitational-wave event.
Some models for prompt radio emission predict a \textit{precursor} signal released before binary merger \citep{Hansen2001,Lyutikov2013,Metzger2016,Lyutikov2018,Wang2018}, and so we additionally analyze the 70 minutes of buffered data recorded before the event.
Our final data set comprises 610 integrations spanning 09:00:03 to 11:12:00 UTC, each 13\,s in duration.

Note that astrophysical signals will also be dispersed \textit{within} each frequency channel.
This intra-channel dispersion is strongest in the lowest channel, in which signals separated by the 24 kHz bandwidth are delayed by a maximum of 6.4\,s with respect to one another.
Although this delay is smaller than our 13\,s integration time, it is sufficiently long that a randomly placed transient might conceivably be split across adjacent integrations, potentially degrading our search sensitivity at low frequencies.

Scatter broadening is unlikely to affect our search.
Assuming a $\nu^{-4}$ frequency dependence, the estimated Milky Way scattering timescale of $0.06\,\mu\mathrm{s}$ at 1\,GHz corresponds only to $0.1\,\mathrm{s}$ at 28\,MHz, much less than our 13\,s integration time \citep{Cordes2002}.
We might expect similarly negligible contributions from GW170104's host galaxy.
Additionally, fast radio bursts show minimal scattering due to the intergalactic medium \citep{Cordes2016}.

Our search window spans approximately 130\,minutes.
In this time the sky rotates considerably, and so we must track the movement of a given source across the OVRO-LWA's field of view.
Just as a sufficiently broadened pulse could span multiple time integrations at a given frequency, it is possible for the Earth's rotation to smear emission across multiple image pixels within a single 13\,s integration.
Since the array's synthesized beam (with 0.50\,deg and 0.24\,deg major and minor axes at 56\,MHz) is larger than our 0.125\,deg pixel size, any astrophysical emission will manifest in multiple neighboring pixels.
We are therefore unlikely to miss a significant fraction of a source's emission as we follow it from one image pixel to the next.

As an example, Fig. \ref{fig:spectrogram} shows the dynamic spectrum obtained by tracking a randomly chosen location within the GW170104 localization region.
To account for slow temporal variations and sidelobes from bright, nearby sources, we have subtracted away the median flux measured in an annulus extending five to seven beamwidths around the target location (see Fig. \ref{fig:meteor} below).
We search all such dynamic spectra for significant dispersed transients, stepping through dispersion measures and times $t_0$ at which a proposed signal enters the OVRO-LWA band.
The spacing $\delta\mathrm{DM}$ between our dispersion measure trials is set by our $t_\mathrm{int}=13\,\mathrm{s}$ integration time and the bounds $\nu_{1}=27.384$\,MHz and $\nu_2=84.912$\,MHz on the OVRO-LWA band:
    \begin{equation}
    \frac{\delta\mathrm{DM}}{\mathrm{pc}\,\mathrm{cm}^{-3}}
        = \frac{t_\mathrm{int}}{4.149\times10^3\,\mathrm{s}}\left[
            \left(\frac{\nu_\mathrm{1}}{\mathrm{MHz}}\right)^{-2}
            -\left(\frac{\nu_\mathrm{2}}{\mathrm{MHz}}\right)^{-2}
        \right]^{-1},
    \end{equation}
giving $\delta\mathrm{DM}=2.62\,\mathrm{pc}\,\mathrm{cm}^{-3}$.
For each dispersed track, we estimate the corresponding flux density with the weighted average (see Appendix \ref{sec:pointEstimates})
    \begin{equation}
    \label{Fhat}
    \hat F = \frac{
        \sum_i \hat F_i/\sigma^2_i
        }{
        \sum_j 1/\sigma^2_j
        },
    \end{equation}
where $\hat F_i$ is the measured flux density in the track's $i$th time-frequency pixel and $\sigma^2_i$ is the corresponding variance, estimated using the background annulus.
In the presence of a true radio transient of flux density $F$, the expectation value and variance of $\hat F$ are
    \begin{equation}
    \langle \hat F \rangle = F
    \end{equation}
and
    \begin{equation}
    \label{sigma2}
    \sigma^2 = \frac{1}{\sum_i 1/\sigma^2_i},
    \end{equation}
respectively.
The signal-to-noise ratio (S/N) of each dispersion trial is defined by combining Eqs. \eqref{Fhat} and \eqref{sigma2}:
    \begin{equation}
    \mathrm{S/N} = \frac{\hat F}{\sigma}.
    \end{equation}
Figure \ref{fig:waterfall}, for example, shows the signal-to-noise ratios obtained from de-dispersing the dynamic spectrum in Fig. \ref{fig:spectrogram}.

\begin{figure}
\centering
\includegraphics[width=0.45\textwidth]{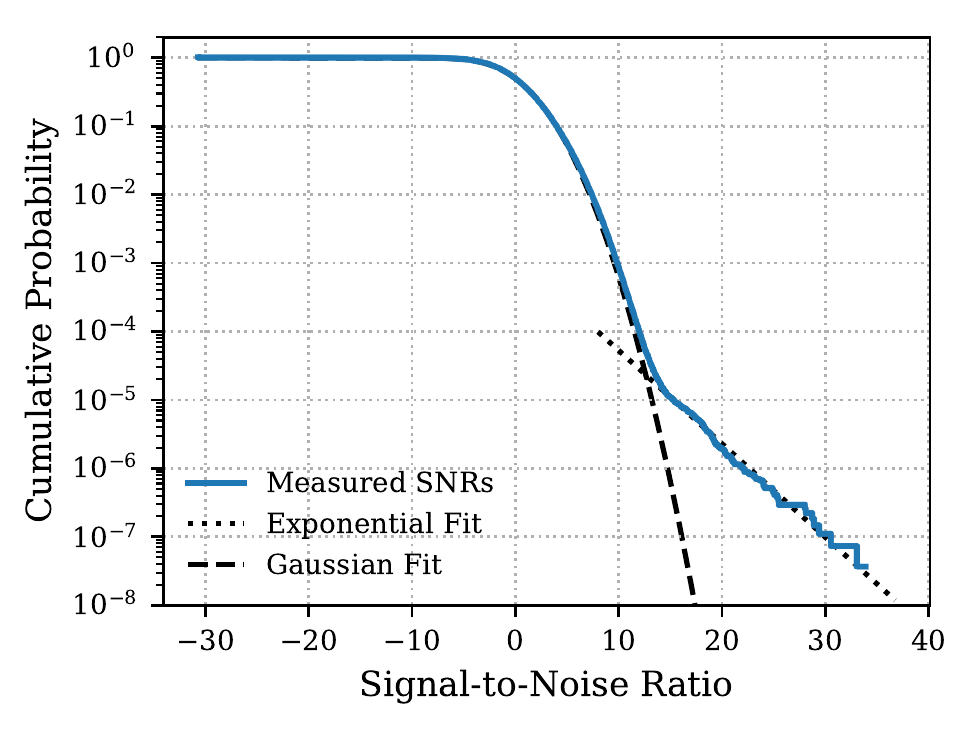}
\caption{
Cumulative background distribution of signal-to-noise ratios from a subset of sky directions and dispersion trials.
The distribution is well fit by a central Gaussian and a exponential tail dominated by meteor reflection events.
Based on this distribution, we manually follow-up any dispersion trial giving $\mathrm{S/N}>20$.
}
\label{fig:background}
\end{figure}

\begin{figure*}
\centering
\includegraphics[height=6cm]{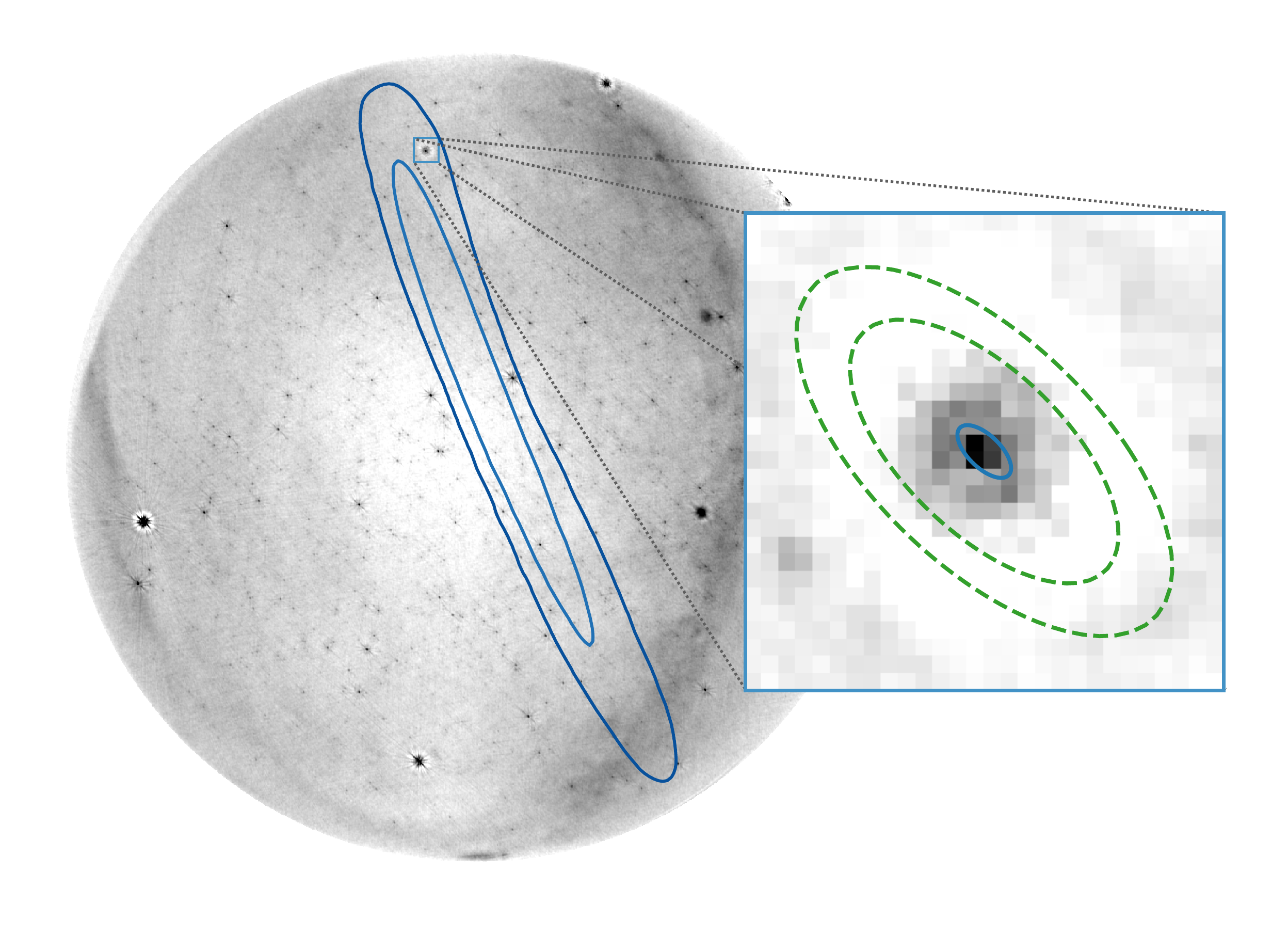}\hspace{0.2cm}
\includegraphics[height=5.6cm]{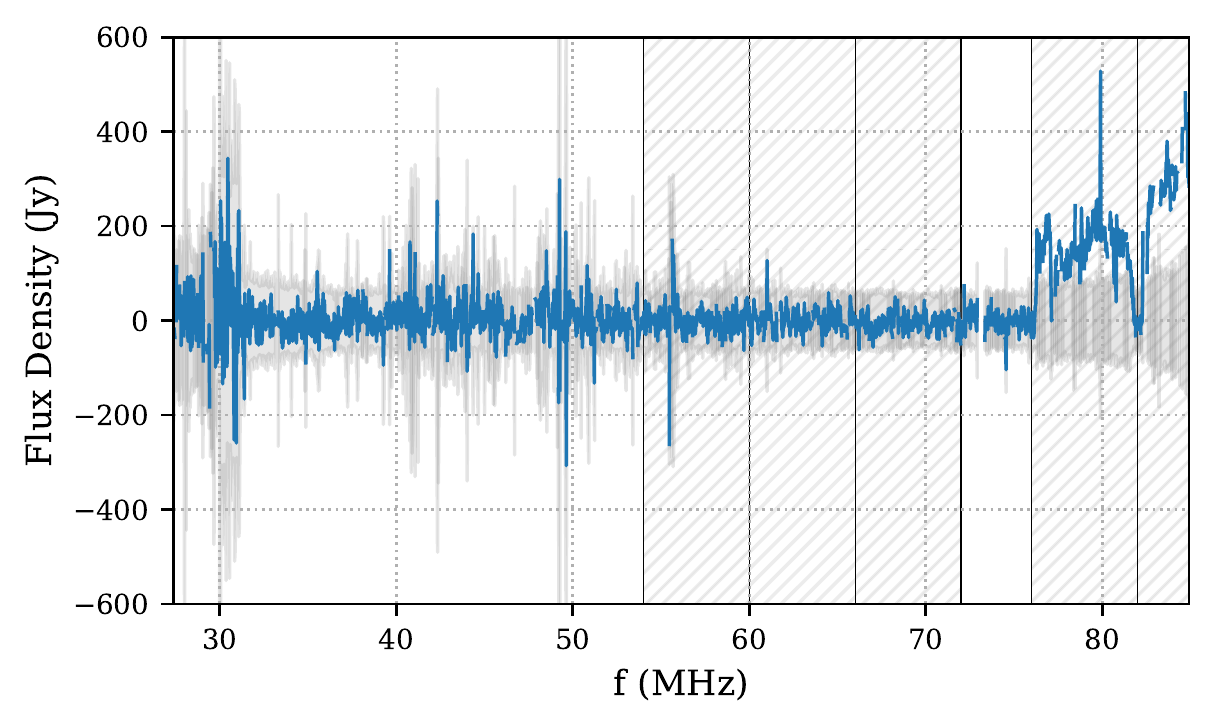}
\caption{
Example of a meteor reflection event.
The left-hand subplot and inset show a full-band dirty image of the reflection; within the inset, the blue contour gives the OVRO-LWA's synthesized beam and the dashed green contours mark the annulus used for background estimation.
Meteor reflections occur in-atmosphere and so appear as resolved sources.
The right-hand subplot shows the spectrum of this event.
As is typical, the observed emission is confined exactly to broadcast televion channels 5 (76--82\,MHz) and 6 (82--88\,MHz), and so is readily identifiable as terrestrial in origin.
}
\label{fig:meteor}
\end{figure*}

With 72,556 sky pixels and 92,763 DM and $t_0$ trials per pixel, a dedispersion search over the entire GW170104 localization region yields $6.7\times10^{9}$ total trials.
To determine a suitable S/N threshold for manual follow-up, in Fig. \ref{fig:background} we plot the cumulative distribution of signal-to-noise ratios obtained from a random subset of sky locations, dispersion measures, and initial times $t_0$.
We find our signal-to-noise ratios to be fairly Gaussian distributed.
The bulk of the distribution is well fit by a somewhat broadened Gaussian centered at zero with a variance of 1.44.
At high significances, however, Fig. \ref{fig:background} shows the emergence of a non-Gaussian tail.
This tail is dominated by meteor reflection events, in which patches of atmosphere temporarily ionized by passing meteors act as reflective surfaces, redirecting RFI from beyond the horizon into the OVRO-LWA (see more below).
The tail is well-fit by $\log_{10}P = a\,(\mathrm{S/N}) + b$, with $a = -0.136$ and $b=-2.913$.
Using this fit, we choose our threshold for manual inspection to be $\mathrm{S/N} = 20$, above which we expect $\left(6.7\times 10^9\right)10^{\,a(\mathrm{S/N})+b} \approx 1.5\times10^4$ outliers.

After searching across the entire GW170104 localization region, we find 6,828 outliers exceeding our threshold.
This suggests that extrapolation of the subset of data shown in Fig. \ref{fig:background} overestimates the rate of high significance events by a factor of two.
All candidates warranting manual follow-up are identified as meteor reflection events \citep{meteors,meteors2}.
Figure \ref{fig:meteor} illustrates the properties of a typical reflection event.
First, meteor reflections occur within the atmosphere (well inside the array's $2 D^2/\lambda \sim 1000\,\mathrm{km}$ far-field limit) and hence appear as resolved sources.
Second, their spectra show emission confined to one or more broadcast television channels.
The reflection event in Fig. \ref{fig:meteor}, for instance, is confined to channels 5 (76--82\,MHz) and 6 (82--88\,MHz).
As meteor reflections currently dominate our search background, the automated identification and rejection of meteor reflections will be a crucial step in improving the sensitivity of future searches.

%An example of one of these reflection events is given in the left-hand side of Fig. \ref{fig:meteor}, which shows the OVRO-LWA sky at the time of a reflection event (09:44:10.1 UTC).
%Within the magnified inset, the blue ellipse denotes the OVRO-LWA's synthsized beam, while the dashed green ellipses bound the annulus used for background subtraction.
%Occurring within the atmosphere, meteor reflections are well within the array's $2 D^2/\lambda \sim 100\,\mathrm{km}$ far-field limit, and so appear as resolved sources.
%The right-hand side of Fig. \ref{fig:meteor}, meanwhile, shows the spectrum of this event.
%Meteor reflections are characterized by emission confined to one or more broadcast television channels -- in this case, channels 5 (76--82\,MHz) and 6 (82--88\,MHz).

%%%%%%%%%%%%%%%%%%%%%%%%%%%%%%%%%
\section{Radio Luminosity Limits}
\label{sec:luminosity}
%%%%%%%%%%%%%%%%%%%%%%%%%%%%%%%%%

\begin{figure*}
\centering
\includegraphics[width=0.7\textwidth]{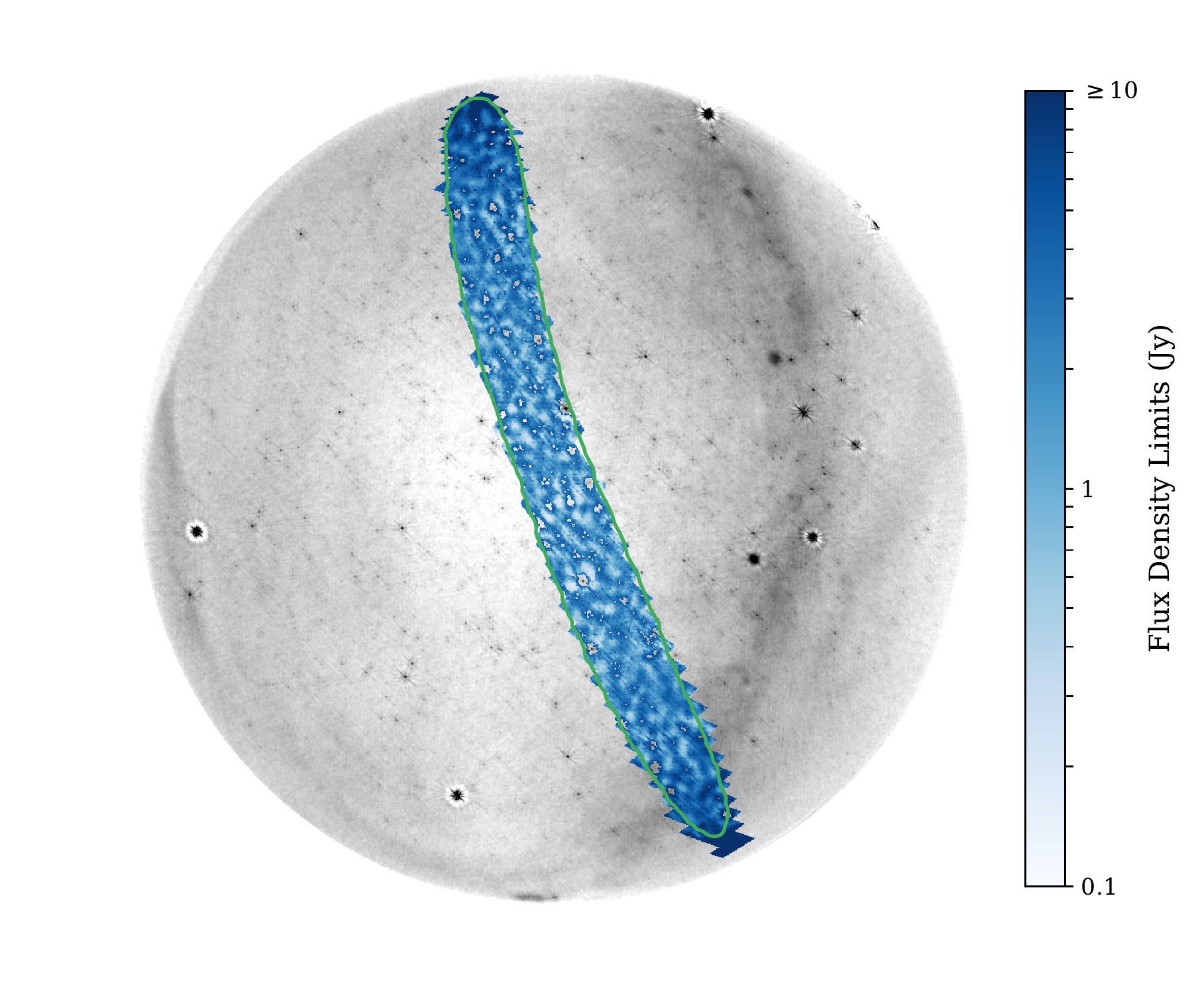}
\caption{
95\% credible upper limits on the flux density of prompt radio emission from GW170104, as a function of its presumed sky location.
The ``holes'' mark locations of persistent point sources excluded from our analysis.
For reference, the contour traces the 95\% credible localization of GW170104 within the OVRO-LWA's field of view.
Our median upper limit across the sky is 2.4 Jy.
Marginalizing over the sky location and distance constraints due to the gravitational-wave signal, we limit GW170104's equivalent isotropic luminosity between 27-84 MHz to $L\leq2.5\times10^{41} \,\mathrm{erg}\,\mathrm{s}^{-1}$ at 95\% credibility.
}
\label{fig:limits}
\end{figure*}

Having rejected all outliers as reflection events, we place upper limits on the prompt radio emission associated with GW170104, following the Bayesian approach described in Appendix \ref{sec:bayes}.
Figure \ref{fig:limits} shows our 95\% credible flux upper limits for each pixel within the 95\% credible gravitational-wave localization region.
We exclude pixels containing persistent point sources detected at $5\sigma$ prior to dedispersion, yielding the ``holes'' seen in Fig. \ref{fig:limits}.
We additionally trim the southernmost points that set below the OVRO-LWA's horizon during the observation.
All together, we cover $94\%$ of the localization region contained within the OVRO-LWA's field of view, and $54\%$ of GW170104's global probability map.
We achieve a median upper limit of 2.4 Jy.
Our sensitivity is degraded at low elevations due to the $(\sin \theta)^{1.6}$ scaling of the antennas' primary beam with elevation angle $\theta$ \citep{Hicks2012}.
Flux upper limits are also impacted by sidelobes in the vicinity of particularly bright point sources.

% Luminosity density = 4.4e33
% times 57.528 MHz = 2.5e41 erg/s

With the sky and distance localization provided by Advanced LIGO, we can re-express our flux limits as constraints on the equivalent isotropic radio luminosity of GW170104.
Marginalizing over the sky location and distance of GW170104 (see Appendix \ref{sec:bayes}), we limit its equivalent isotropic luminosity between 27-84 MHz to $L_\mathrm{radio}\leq 2.5\times10^{41}\,\mathrm{erg}\,\mathrm{s}^{-1}$ at 95\% credibility, assuming the source lies within the OVRO-LWA's field of view.
The total energy radiated by GW170104 was $E_\mathrm{GW} = 2.0^{+0.6}_{-0.7} M_\odot c^2$ \citep{gw170104}.
We therefore limit the fraction $f$ of the total energy converted to prompt radio emission to be $f= L_\mathrm{radio} t_\mathrm{int}/E_\mathrm{GW} \leq 1.4\times10^{-12}$, using the lower bound on $E_\mathrm{GW}$.

Meanwhile, the equivalent luminosity of the \textit{Fermi-GBM} outlier associated with GW150914 was $L_\gamma = 1.8^{+1.5}_{-1.0}\times10^{49}\,\mathrm{erg}\,\mathrm{s}^{-1}$ \citep{fermi_2016}.
Note that GW150914 and GW170104 occurred at luminosity distances of approximately $410\,\mathrm{Mpc}$ and $880\,\mathrm{Mpc}$, respectively \citep{GW150914,gw170104}.
If the OVRO-LWA had been operating at the time of GW150914, we would therefore have been sensitive to any associated radio transient with luminosity $L_\mathrm{radio} \approx 2.5\times10^{41}\,\mathrm{erg}\,\mathrm{s}^{-1} \left(410\,\mathrm{Mpc}/880\,\mathrm{Mpc}\right)^2 \approx 5.5\times 10^{40}\,\mathrm{erg}\,\mathrm{s}^{-1}$.
If, in the future, additional gamma-ray outliers are identified in coincidence with gravitational-wave events, simultaneous observation with the OVRO-LWA will enable limits on the relative radio radio and gamma-ray emission of order $L_\mathrm{radio}/L_\gamma \lesssim 3\times10^{-9}$.

%%%%%%%%%%%%%%%%%%%%%%%%%%%%%%%%%
\section{The Third LIGO/Virgo Observing Run and Beyond}
\label{sec:future}
%%%%%%%%%%%%%%%%%%%%%%%%%%%%%%%%%

Advanced LIGO \& Virgo's third observing run (O3) is scheduled to commence in April 2019 and run for one calendar year.
During this time, between 1-50 binary neutron star detections are expected \citep{Prospects}.
The OVRO-LWA will operate in continuous buffering mode during O3, searching for prompt radio transients associated with compact binary mergers.

The sensitivity of this study to sub-second radio transients is limited by the 13\,s resolution of buffered visibilities.
The buffering of future data with higher time resolution will increase the signal-to-noise ratio of temporally unresolved transients.
We are additionally exploring options to buffer the incoherent sum of antenna powers at their raw 197\,MHz sampling rate.
The incoherent sum will provide no directional information, but the vastly increased time resolution and temporal coincidence with gravitational-wave events will enable sensitive measurements of prompt radio transients.

A more ambitious goal is the buffering and coherent dedispersion of all 512 signal paths at 197\,MHz.
This endeavor has previously required prohibitively large buffer disk space due to significant latency in the release of LIGO/Virgo alerts.
In their upcoming O3 observing run, however, LIGO \& Virgo will transition to automated alerts released within 1-10 minutes of a gravitational-wave candidate \citep{emfollowWebsite}.
If successful, this reduced latency may make the buffering of raw antenna voltages computationally feasible.

Finally, the OVRO-LWA will soon be undergoing upgrades towards its ``Stage 3" design, consisting of 352 correlated antennas over an extended 2.5\,km maximum baseline.
Also included in this design is the buffering of raw antenna voltages, allowing high time-resolution searches triggered by automated LIGO \& Virgo alerts.
With these improvements, Stage 3 OVRO-LWA promises to enable even more sensitive detection and precise localization of prompt radio emission from compact binary mergers. 

\acknowledgments
This material is based in part upon work supported by the National Science Foundation under Grant AST-1654815 and AST-1212226.
TC is supported by LIGO Laboratory, funded by the National Science Foundation under cooperative agreement PHY-0757058, and by the Josephine de Karman Fellowship Trust.
GH acknowledges the support of the Alfred P. Sloan Foundation and the Research Corporation for Science Advancement.
The OVRO-LWA project was initiated through the kind donation of Deborah Castleman and Harold Rosen.
% cooperative agreement PHY-0757058

\bibliography{References.bib}

\begin{thebibliography}{}
\expandafter\ifx\csname natexlab\endcsname\relax\def\natexlab#1{#1}\fi
\providecommand{\url}[1]{\href{#1}{#1}}

\bibitem[{Abbott {et~al.}(2016)Abbott, Abbott, Abbott, Abernathy, Acernese,
  {et~al.}}]{GW150914}
Abbott, B.~P., Abbott, R., Abbott, T.~D., {et~al.} 2016, Phys. Rev. Lett., 116,
  061102.
\newblock \url{http://link.aps.org/doi/10.1103/PhysRevLett.116.061102}

\bibitem[{Abbott {et~al.}(2018{\natexlab{a}})Abbott, Abbott, Abbott, Abernathy,
  Acernese, {et~al.}}]{Prospects}
---. 2018{\natexlab{a}}, Living Reviews in Relativity, 21, 3.
\newblock \url{https://doi.org/10.1007/s41114-018-0012-9}

\bibitem[{Abbott {et~al.}(2019)Abbott, Abbott, Abbott, Abraham, Acernese,
  {et~al.}}]{LigoLowLatency}
---. 2019, arXiv:1901.03310.
\newblock \url{http://arxiv.org/abs/1901.03310}

\bibitem[{Abbott {et~al.}(2017{\natexlab{a}})Abbott, Abbott, Abbott, Acernese,
  Ackley, {et~al.}}]{BNS_gw}
---. 2017{\natexlab{a}}, Phys. Rev. Lett., 119, 161101.
\newblock \url{https://link.aps.org/doi/10.1103/PhysRevLett.119.161101}

\bibitem[{Abbott {et~al.}(2017{\natexlab{b}})Abbott, Abbott, Abbott, Acernese,
  Ackley, {et~al.}}]{BNS_mma}
---. 2017{\natexlab{b}}, Astrophys. J., 848, L12.
\newblock \url{http://iopscience.iop.org/article/10.3847/2041-8213/aa91c9}

\bibitem[{Abbott {et~al.}(2017{\natexlab{c}})Abbott, Abbott, Abbott, Acernese,
  Ackley, {et~al.}}]{BNS_H0}
---. 2017{\natexlab{c}}, Nature, 551, 85.
\newblock \url{http://dx.doi.org/10.1038/nature24471}

\bibitem[{Abbott {et~al.}(2017{\natexlab{d}})Abbott, Abbott, Abbott, Acernese,
  Ackley, {et~al.}}]{gw170104}
---. 2017{\natexlab{d}}, Phys. Rev. Lett., 118, 221101.
\newblock \url{http://link.aps.org/doi/10.1103/PhysRevLett.118.221101}

\bibitem[{Abbott {et~al.}(2017{\natexlab{e}})Abbott, Abbott, Abbott, Acernese,
  Ackley, {et~al.}}]{gw170104_supplement}
---. 2017{\natexlab{e}}, Phys. Rev. Lett., 118, 221101 (Supplement).
\newblock \url{http://link.aps.org/doi/10.1103/PhysRevLett.118.221101}

\bibitem[{Abbott {et~al.}(2018{\natexlab{b}})Abbott, Abbott, Abbott, Acernese,
  {Ackley}, {et~al.}}]{BNS_eos2}
---. 2018{\natexlab{b}}, Phys. Rev. Lett., 121, 161101.
\newblock \url{http://doi.org/10.1103/PhysRevLett.121.161101}

\bibitem[{Anderson {et~al.}(2018)Anderson, Hallinan, Eastwood, Monroe,
  Vedantham, {et~al.}}]{Anderson2018}
Anderson, M.~M., Hallinan, G., Eastwood, M.~W., {et~al.} 2018, Astrophys. J.,
  864, 22.
\newblock \url{https://doi.org/10.3847/1538-4357/aad2d7}

\bibitem[{Baars {et~al.}(1977)Baars, Genzel, Pauliny-Toth, \&
  Witzel}]{Baars1977}
Baars, J. W.~M., Genzel, R., Pauliny-Toth, I. I.~K., \& Witzel, A. 1977,
  Astron. Astrophys., 61, 99.
\newblock \url{http://adsabs.harvard.edu/abs/1977A\%26A....61...99B}

\bibitem[{Bannister {et~al.}(2012)Bannister, Murphy, Gaensler, \&
  Reynolds}]{Bannister2012}
Bannister, K.~W., Murphy, T., Gaensler, B.~M., \& Reynolds, J.~E. 2012,
  Astrophys. J., 757.
\newblock \url{http://doi.org/10.1088/0004-637X/757/1/38}

\bibitem[{Berger(2014)}]{Berger2014}
Berger, E. 2014, Ann. Rev. Astron. and Astrophys., 52, 43.
\newblock \url{https://doi.org/10.1146/annurev-astro-081913-035926}

\bibitem[{Callister {et~al.}(2017)Callister, Dowell, Kanner,
  {et~al.}}]{LWA-170817}
Callister, T., Dowell, J., Kanner, J., {et~al.} 2017, GCN, 21680

\bibitem[{{Ceplecha} {et~al.}(1998){Ceplecha}, {Borovi{\v c}ka}, {Elford},
  {Revelle}, {Hawkes}, {Porub{\v c}an}, \& {{\v S}imek}}]{meteors}
{Ceplecha}, Z., {Borovi{\v c}ka}, J., {Elford}, W.~G., {et~al.} 1998, Space
  Science Rev., 84, 327.
\newblock \url{http://doi.org/10.1023/A:1005069928850}

\bibitem[{Chu {et~al.}(2016)Chu, Howell, Rowlinson, Gao, Zhang, Tingay,
  Bo{\"{e}}r, \& Wen}]{ChuProspects}
Chu, Q., Howell, E.~J., Rowlinson, A., {et~al.} 2016, Mon. Not. R. Astron.
  Soc., 459, 121.
\newblock \url{http://doi.org/10.1093/mnras/stw576}

\bibitem[{Connaughton {et~al.}(2016)Connaughton, Burns, Goldstein, Blackburn,
  Briggs, Zhang, Camp, Christensen, Hui, Jenke, Littenberg, McEnery, Racusin,
  Shawhan, Singer, Veitch, Wilson-Hodge, Bhat, Bissaldi, Cleveland,
  Fitzpatrick, Giles, Gibby, von Kienlin, Kippen, McBreen, Mailyan, Meegan,
  Paciesas, Preece, Roberts, Sparke, Stanbro, Toelge, \& Veres}]{fermi_2016}
Connaughton, V., Burns, E., Goldstein, A., {et~al.} 2016, Astrophys. J., 826,
  L6.
\newblock \url{http://doi.org/10.3847/2041-8205/826/1/L6}

\bibitem[{Connaughton {et~al.}(2018)Connaughton, Burns, Goldstein, Blackburn,
  Briggs, Christensen, Hui, Kocevski, Littenberg, McEnery, Racusin, Shawhan,
  Veitch, Wilson-Hodge, Bhat, Bissaldi, Cleveland, Giles, Gibby, Kienlin,
  Kippen, McBreen, Meegan, Paciesas, Preece, Roberts, Stanbro, \&
  Veres}]{Connaughton2018}
---. 2018, Astrophys. J., 853, L9.
\newblock \url{http://doi.org/10.3847/2041-8213/aaa4f2}

\bibitem[{Cordes \& Lazio(2002)}]{Cordes2002}
Cordes, J.~M., \& Lazio, T. J.~W. 2002, arXiv: astro-ph/0207156.
\newblock \url{http://arxiv.org/abs/astro-ph/0207156}

\bibitem[{Cordes \& Lazio(2003)}]{Cordes2003}
---. 2003, arXiv: astro-ph/0301598.
\newblock \url{http://arxiv.org/abs/astro-ph/0301598}

\bibitem[{Cordes {et~al.}(2016)Cordes, Wharton, Spitler, Chatterjee, \&
  Wasserman}]{Cordes2016}
Cordes, J.~M., Wharton, R.~S., Spitler, L.~G., Chatterjee, S., \& Wasserman, I.
  2016, arXiv: 1605.05890.
\newblock \url{http://arxiv.org/abs/1605.05890}

\bibitem[{Coulter {et~al.}(2017)Coulter, Foley, Kilpatrick, Drout, Piro,
  {et~al.}}]{BNS_optical}
Coulter, D.~A., Foley, R.~J., Kilpatrick, C.~D., {et~al.} 2017, Science, 358,
  1556.
\newblock \url{http://doi.org/10.1126/science.aap9811}

\bibitem[{de~Mink \& King(2017)}]{deMink2017}
de~Mink, S.~E., \& King, A. 2017, Astrophys. J., 839, L7.
\newblock \url{http://doi.org/10.3847/2041-8213/aa67f3}

\bibitem[{Drout {et~al.}(2017)Drout, Piro, Shappee, Kilpatrick, Simon,
  {et~al.}}]{BNS_kilonova1}
Drout, M.~R., Piro, A.~L., Shappee, B.~J., {et~al.} 2017, Science, 358, 1570.
\newblock \url{http://doi.org/10.1126/science.aaq0049}

\bibitem[{Eastwood(2016)}]{ttcal}
Eastwood, M.~W. 2016, {TTCal},  Zenodo, doi:10.5281/zenodo.1049160.
\newblock \url{https://doi.org/10.5281/zenodo.1049160}

\bibitem[{Eastwood {et~al.}(2018)Eastwood, Anderson, Monroe, Hallinan,
  Barsdell, {et~al.}}]{Eastwood2018}
Eastwood, M.~W., Anderson, M.~M., Monroe, R.~M., {et~al.} 2018, Astron. J.,
  156, 32.
\newblock \url{Ahttp://dx.doi.org/10.3847/1538-3881/aac721}

\bibitem[{Fraschetti(2018)}]{Fraschetti2018}
Fraschetti, F. 2018, Journal of Cosmology and Astroparticle Physics, 2018, 054.
\newblock \url{https://doi.org/10.1088/1475-7516/2018/04/054}

\bibitem[{Goldstein {et~al.}(2017)Goldstein, Veres, Burns, Briggs, Hamburg,
  {et~al.}}]{BNS_gbm}
Goldstein, A., Veres, P., Burns, E., {et~al.} 2017, Astrophys. J., 848, L14.
\newblock \url{http://doi.org/10.3847/2041-8213/aa8f41}

\bibitem[{Greiner {et~al.}(2016)Greiner, Burgess, Savchenko, \&
  Yu}]{Greiner2016}
Greiner, J., Burgess, J.~M., Savchenko, V., \& Yu, H.-F. 2016, Astrophys. J.,
  827, L38.
\newblock \url{http://doi.org/10.3847/2041-8205/827/2/L38}

\bibitem[{Hallinan {et~al.}(2017)Hallinan, Corsi, Mooley, Hotokezaka, Nakar,
  {et~al.}}]{BNS_radio}
Hallinan, G., Corsi, A., Mooley, K.~P., {et~al.} 2017, Science, 358, 1579.
\newblock \url{http://www.sciencemag.org/lookup/doi/10.1126/science.aap9855}

\bibitem[{Hansen \& Lyutikov(2001)}]{Hansen2001}
Hansen, B. M.~S., \& Lyutikov, M. 2001, Mon. Not. R. Astron. Soc., 322, 695.
\newblock \url{http://doi.org/10.1046/j.1365-8711.2001.04103.x}

\bibitem[{Helmboldt {et~al.}(2014)Helmboldt, Ellingson, Hartman, Lazio, Taylor,
  Wilson, \& Wolfe}]{meteors2}
Helmboldt, J.~F., Ellingson, S.~W., Hartman, J.~M., {et~al.} 2014, Radio
  Science, 49, 157.
\newblock
  \url{https://agupubs.onlinelibrary.wiley.com/doi/abs/10.1002/2013RS005220}

\bibitem[{Hicks {et~al.}(2012)Hicks, Paravastu-Dalal, Stewart, Erickson, Ray,
  Kassim, Burns, Clarke, Schmitt, Craig, Hartman, \& Weiler}]{Hicks2012}
Hicks, B.~C., Paravastu-Dalal, N., Stewart, K.~P., {et~al.} 2012, Publications
  of the Astronomical Society of the Pacific, 124, 1090.
\newblock \url{http://iopscience.iop.org/article/10.1086/668121}

\bibitem[{{Hotokezaka} {et~al.}(2018){Hotokezaka}, {Nakar}, {Gottlieb},
  {Nissanke}, {Masuda}, {Hallinan}, {Mooley}, \& {Deller}}]{BNS_H02}
{Hotokezaka}, K., {Nakar}, E., {Gottlieb}, O., {et~al.} 2018, arXiv:1806.10596.
\newblock \url{http://arxiv.org/abs/1806.10596}

\bibitem[{{Inoue}(2004)}]{Inoue2004}
{Inoue}, S. 2004, Mon. Not. R. Astron. Soc., 348, 999.
\newblock \url{http://doi.org/10.1111/j.1365-2966.2004.07359.x}

\bibitem[{{Ioka}(2003)}]{Ioka2003}
{Ioka}, K. 2003, Astrophys. J. Lett., 598, L79.
\newblock \url{http://doi.org/10.1086/380598}

\bibitem[{Kaplan {et~al.}(2017)Kaplan, Brown, Sokolowski, Wayth,
  {et~al.}}]{MWA-170817}
Kaplan, D., Brown, I., Sokolowski, M., Wayth, R., {et~al.} 2017, GCN, 21927

\bibitem[{Kaplan {et~al.}(2016)Kaplan, Murphy, Rowlinson, Croft, Wayth, \&
  Trott}]{KaplanProspects}
Kaplan, D.~L., Murphy, T., Rowlinson, A., {et~al.} 2016, Publications of the
  Astronomical Society of Australia, 33, e050.
\newblock \url{http://doi.org/10.1017/pasa.2016.43}

\bibitem[{Kaplan {et~al.}(2015)Kaplan, Rowlinson, Bannister, Bell, Croft,
  Murphy, Tingay, Wayth, \& Williams}]{Kaplan2015}
Kaplan, D.~L., Rowlinson, A., Bannister, K.~W., {et~al.} 2015, Astrophys. J.,
  814, L25.
\newblock \url{http://doi.org/10.1088/2041-8205/814/2/L25}

\bibitem[{{Kocz} {et~al.}(2015){Kocz}, {Greenhill}, {Barsdell}, {Price},
  {Bernardi}, {et~al.}}]{leda}
{Kocz}, J., {Greenhill}, L.~J., {Barsdell}, B.~R., {et~al.} 2015, Journal of
  Astronomical Instrumentation, 4, 1550003.
\newblock \url{http://doi.org/10.1142/S2251171715500038}

\bibitem[{Lai(2012)}]{Lai2012}
Lai, D. 2012, Astrophys. J. Lett., 757, 1.
\newblock \url{http://doi.org/10.1088/2041-8205/757/1/L3}

\bibitem[{Liebling \& Palenzuela(2016)}]{Liebling2016}
Liebling, S.~L., \& Palenzuela, C. 2016, Phys. Rev. D, 94,
  doi:10.1103/PhysRevD.94.064046.
\newblock \url{http://doi.org/10.1103/PhysRevD.94.064046}

\bibitem[{{LIGO Scientific Collaboration \& Virgo
  Collaboration}(2018)}]{emfollowWebsite}
{LIGO Scientific Collaboration \& Virgo Collaboration}. 2018,
  \url{https://emfollow.docs.ligo.org/userguide/index.html}, {Accessed:
  2019-02-25}

\bibitem[{Liu {et~al.}(2016)Liu, Romero, Liu, \& Li}]{Liu2016}
Liu, T., Romero, G.~E., Liu, M.-L., \& Li, A. 2016, Astrophys. J., 826, 82.
\newblock \url{http://doi.org/10.3847/0004-637X/826/1/82}

\bibitem[{Loeb(2016)}]{Loeb2016}
Loeb, A. 2016, Astrophys. J., 819, L21.
\newblock \url{http://doi.org/10.3847/2041-8205/819/2/L21}

\bibitem[{Lyutikov(2013)}]{Lyutikov2013}
Lyutikov, M. 2013, Astrophys. J., 768, 63.
\newblock \url{https://doi.org/10.1088/0004-637X/768/1/63}

\bibitem[{Lyutikov(2016)}]{LyutikovFermi2016}
---. 2016, arXiv: 1602.07352.
\newblock \url{http://arxiv.org/abs/1602.07352}

\bibitem[{Lyutikov(2018)}]{Lyutikov2018}
---. 2018, arXiv:1809.10478, arXiv:1809.10478.
\newblock \url{http://arxiv.org/abs/1809.10478}

\bibitem[{Metzger \& Zivancev(2016)}]{Metzger2016}
Metzger, B.~D., \& Zivancev, C. 2016, Mon. Not. R. Astron. Soc., 461, 4435.
\newblock \url{https://doi.org/10.1093/mnras/stw1800}

\bibitem[{{Mooley} {et~al.}(2018){Mooley}, {Frail}, {Dobie}, {Lenc}, {Corsi},
  {et~al.}}]{BNS_jet2}
{Mooley}, K.~P., {Frail}, D.~A., {Dobie}, D., {et~al.} 2018, Astrophys. J.,
  868, L11.
\newblock \url{http://doi.org/10.3847/2041-8213/aaeda7}

\bibitem[{{Nakar} {et~al.}(2018){Nakar}, {Gottlieb}, {Piran}, {Kasliwal}, \&
  {Hallinan}}]{BNS_jet1}
{Nakar}, E., {Gottlieb}, O., {Piran}, T., {Kasliwal}, M.~M., \& {Hallinan}, G.
  2018, Astrophys. J., 867, 18.
\newblock \url{http://doi.org/10.3847/1538-4357/aae205}

\bibitem[{Obenberger {et~al.}(2014)Obenberger, Hartman, Taylor, Craig, Dowell,
  Helmboldt, Henning, Schinzel, \& Wilson}]{Obenberger2014}
Obenberger, K.~S., Hartman, J.~M., Taylor, G.~B., {et~al.} 2014, Astrophys. J.,
  785, 27.
\newblock \url{http://doi.org/10.1088/0004-637X/785/1/27}

\bibitem[{Offringa {et~al.}(2014)Offringa, McKinley, Hurley-Walker, Briggs,
  Wayth, Kaplan, Bell, Feng, Neben, Hughes, Rhee, Murphy, Bhat, Bernardi,
  Bowman, Cappallo, Corey, Deshpande, Emrich, Ewall-Wice, Gaensler, Goeke,
  Greenhill, Hazelton, Hindson, Johnston-Hollitt, Jacobs, Kasper, Kratzenberg,
  Lenc, Lonsdale, Lynch, McWhirter, Mitchell, Morales, Morgan, Kudryavtseva,
  Oberoi, Ord, Pindor, Procopio, Prabu, Riding, Roshi, Shankar, Srivani,
  Subrahmanyan, Tingay, Waterson, Webster, Whitney, Williams, \&
  Williams}]{Offringa2014}
Offringa, A.~R., McKinley, B., Hurley-Walker, N., {et~al.} 2014, Mon. Not. R.
  Astron. Soc., 444, 606.
\newblock \url{http://doi.org/10.1093/mnras/stu1368}

\bibitem[{Perley \& Butler(2017)}]{Perley2017}
Perley, R.~A., \& Butler, B.~J. 2017, Astrophys. J. Supplement Series, 230, 7.
\newblock \url{http://doi.org/10.3847/1538-4365/aa6df9}

\bibitem[{Perna {et~al.}(2016)Perna, Lazzati, \& Giacomazzo}]{Perna2016}
Perna, R., Lazzati, D., \& Giacomazzo, B. 2016, Astrophys. J., 821, L18.
\newblock \url{http://doi.org/10.3847/2041-8205/821/1/L18}

\bibitem[{Pshirkov \& Postnov(2010)}]{Pshirkov2010}
Pshirkov, M.~S., \& Postnov, K.~A. 2010, Astrophys. Space Sci., 330, 13.
\newblock \url{http://link.springer.com/10.1007/s10509-010-0395-x}

\bibitem[{{Raithel} {et~al.}(2018){Raithel}, {{\"O}zel}, \&
  {Psaltis}}]{BNS_eos1}
{Raithel}, C.~A., {{\"O}zel}, F., \& {Psaltis}, D. 2018, Astrophys. J., 857,
  L23.
\newblock \url{http://doi.org/10.3847/2041-8213/aabcbf}

\bibitem[{Ravi \& Lasky(2014)}]{Ravi2014}
Ravi, V., \& Lasky, P.~D. 2014, Mon. Not. R. Astron. Soc., 441, 2433.
\newblock \url{http://doi.org/10.1093/mnras/stu720}

\bibitem[{Rodriguez {et~al.}(2018)Rodriguez, Amaro-Seoane, Chatterjee, Kremer,
  Rasio, Samsing, Ye, \& Zevin}]{Rodriguez2018}
Rodriguez, C.~L., Amaro-Seoane, P., Chatterjee, S., {et~al.} 2018, Phys. Rev.
  D, 98, 123005.
\newblock \url{https://link.aps.org/doi/10.1103/PhysRevD.98.123005}

\bibitem[{Savchenko {et~al.}(2016)Savchenko, Ferrigno, Mereghetti, Natalucci,
  Bazzano, Bozzo, Brandt, Courvoisier, Diehl, Hanlon, Kienlin, Kuulkers,
  Laurent, Lebrun, Roques, Ubertini, \& Weidenspointner}]{Savchenko2016}
Savchenko, V., Ferrigno, C., Mereghetti, S., {et~al.} 2016, Astrophys. J., 820,
  L36.
\newblock \url{http://doi.org/10.3847/2041-8205/820/2/L36}

\bibitem[{Smartt {et~al.}(2017)Smartt, Chen, Jerkstrand, Coughlin, Kankare,
  {et~al.}}]{BNS_kilonova2}
Smartt, S.~J., Chen, T.-W., Jerkstrand, A., {et~al.} 2017, Nature, 551, 75.
\newblock \url{http://dx.doi.org/10.1038/nature24303}

\bibitem[{{Totani}(2013)}]{Totani2013}
{Totani}, T. 2013, Pub. Astron. Soc. Japan, 65, L12.
\newblock \url{https://doi.org/10.1093/pasj/65.5.L12}

\bibitem[{Troja {et~al.}(2017)Troja, Piro, van Eerten, Wollaeger, Im,
  {et~al.}}]{BNS_xray}
Troja, E., Piro, L., van Eerten, H., {et~al.} 2017, Nature, 551, 71.
\newblock \url{http://www.nature.com/doifinder/10.1038/nature24290}

\bibitem[{Usov \& Katz(2000)}]{Usov2000}
Usov, V.~V., \& Katz, J.~I. 2000, Astron. Astrophys., 364, 655.
\newblock \url{http://arxiv.org/abs/astro-ph/0002278}

\bibitem[{Wang {et~al.}(2018)Wang, Peng, Wu, \& Dai}]{Wang2018}
Wang, J.-S., Peng, F.-K., Wu, K., \& Dai, Z.-G. 2018, Astrophys. J., 868, 19.
\newblock \url{http://stacks.iop.org/0004-637X/868/i=1/a=19}

\bibitem[{Wang {et~al.}(2016)Wang, Yang, Wu, Dai, \& Wang}]{Wang2016}
Wang, J.-S., Yang, Y.-P., Wu, X.-F., Dai, Z.-G., \& Wang, F.-Y. 2016,
  Astrophys. J. Lett., 822, L7.
\newblock \url{http://stacks.iop.org/2041-8205/822/i=1/a=L7}

\bibitem[{Yancey {et~al.}(2015)Yancey, Bear, Akukwe, Chen, Dowell, Gough,
  Kanner, Kavic, Obenberger, Shawhan, Simonetti, Taylor, \&
  Tsai}]{YanceyProspects}
Yancey, C.~C., Bear, B.~E., Akukwe, B., {et~al.} 2015, Astrophys. J., 812, 168.
\newblock \url{http://dx.doi.org/10.1088/0004-637X/812/2/168}

\bibitem[{Zhang(2016)}]{Zhang2016}
Zhang, B. 2016, Astrophys. J., 827, L31.
\newblock \url{http://doi.org/10.3847/2041-8205/827/2/L31}

\end{thebibliography}

%%TC:ignore
%%%%%%%%%%%%%%%%%%%%%%%%%%%%%%%%%%%%%
\appendix
\section{Dispersion Measure Bounds}
\label{sec:dm}
%%%%%%%%%%%%%%%%%%%%%%%%%%%%%%%%%%%%%

Parameter estimation on the gravitational-wave signal GW170104 constrains its redshift to $z=0.173^{+0.072}_{-0.071}$ with an effective-precession waveform model and $z=0.182^{+0.081}_{-0.078}$ using a model capturing full spin-precession effects \citep{gw170104,gw170104_supplement}.
We conservatively assume that GW170104's progenitor lies between $0.1\leq z \leq 0.3$.

Dispersion measure is defined as the integrated column density of free electrons (number density $n_e$) along a given line of sight: $\mathrm{DM} = \int n_e ds$.
When allowing for cosmology, the dispersion measure due to propagation through the intergalactic medium is \citep{Ioka2003,Inoue2004}
    \begin{equation}
    \mathrm{DM}_\textsc{igm} = c \int_0^z \frac{n_e \left(1+z'\right)}{H(z')}dz',
    \end{equation}
where $H(z) = H_0\sqrt{\Omega_m(1+z)^3+\Omega_\Lambda}$, $H_0$ is the Hubble constant, and $\Omega_m$ and $\Omega_\Lambda$ are the dimensionless energy-densities of matter and dark energy, respectively.
We take $H_0 = 67.7\,\mathrm{km}\,\mathrm{s}^{-1}\,\mathrm{Mpc}^{-1}$, $\Omega_m = 0.31$, and $\Omega_\Lambda=0.69$.
An upper limit on mean electron density in the intergalactic medium is obtained by assuming the Universe's baryonic density $\Omega_B = 0.049$ is composed entirely of ionized hydrogen \citep{Ioka2003}.
Then the mean electron number density is $\overline n_e = \Omega_B \rho_c/m_p$, where $\rho_c = 3 H_0^2/8 \pi G$ is the closure density of the Universe, $G$ is Newton's constant, and $m_p$ is the proton mass.
As the Universe is neither fully ionized nor composed purely of hydrogen, this approximation yields an overestimate of $\overline n_e$ and hence a conservative overestimate of the intergalactic dispersion measure.
Assuming that GW170104's progenitor lies within $0.1\leq z \leq 0.3$, we estimate $113\,\mathrm{pc}\,\mathrm{cm}^{-3} \leq \mathrm{DM}_\textsc{igm} \leq 350\,\mathrm{pc}\,\mathrm{cm}^{-3}$.

In addition, the Milky Way, the progenitor's host galaxy, and the progenitor's immediate environment will contribute to the net dispersion measure.
The GW170104 localization region spans a broad range of Galactic latitudes, corresponding to a wide range of possible Galactic dispersion measures.
A lower bound on the Milky Way dispersion measure is simply zero.
An upper bound is given by assuming a line of sight directly through the Galactic disk, yielding $0 \leq\mathrm{DM}_\textsc{mw}\leq 180\,\mathrm{pc}\,\mathrm{cm}^{-3}$ \citep{Cordes2002,Cordes2003}.
We have no knowledge of the progenitor's host galaxy or its environment, and so we naively assume $\mathrm{DM}_\mathrm{Host}+\mathrm{DM}_\mathrm{Env}\leq 100\,\mathrm{pc}\,\mathrm{cm}^{-3}$.
Compact binaries are expected to be spatially offset from their host galaxy due to natal supernova kicks or dynamical ejection from dense clusters \citep{Berger2014,Rodriguez2018}, and so in practice the dispersion measure from the binary's immediate environment may dominate over that from its host.

Combining contributions from the Milky Way, the intergalactic medium, and GW170104's environment and host galaxy, we bound the dispersion measure of radio transients associated with GW170104 to $113\,\mathrm{pc}\,\mathrm{cm}^{-3} \leq \mathrm{DM}_\textsc{igm} \leq 630\,\mathrm{pc}\,\mathrm{cm}^{-3}$.

%%%%%%%%%%%%%%%%%%%%%%%%%%%%%%%%%%
\section{Optimal Flux Density Estimator}
\label{sec:pointEstimates}
%%%%%%%%%%%%%%%%%%%%%%%%%%%%%%%%%%

In Sec. \ref{sec:dedispersion}, we define an estimator
    \begin{equation}
    \label{eq:AppendixFhat}
    \hat F = \frac{
        \sum_i \hat F_i/\sigma^2_i
        }{
        \sum_j 1/\sigma^2_j
        },
    \end{equation}
for the flux density $F$ in a dispersed time-frequency track (indexed by $i$).
This estimator is unbiased (with expectation value $\langle \hat F \rangle = F$) and optimal, maximizing the expected signal-to-noise ratio in the presence of Gaussian noise.

To better understand this, consider estimating the flux density along a time-frequency track with individual flux measurements $\hat F_i$.
This estimate will take the general form of a weighted average,
    \begin{equation}
    \label{eq:arbitraryWeights}
    \hat F = \frac{\sum_i \hat F_i w_i}{\sum_j w_j},
    \end{equation}
where the $w_i$'s are yet-undetermined weights.
If the expected flux in each time-frequency pixel is $\langle \hat F_i \rangle = F$, then the expectation value of $\hat F$ is
    \begin{equation}
    \label{eq:weightedMean}
    \langle \hat F \rangle
        = \frac{\sum_i \langle\hat F_i\rangle w_i}{\sum_j w_j}
        = F.
    \end{equation}
Its variance, meanwhile, is
    \begin{equation}
    \begin{aligned}
    \sigma^2
        &= \langle \hat F^2 \rangle - \langle \hat F \rangle^2 \\
        &= \frac{1}{\left(\sum_k w_k\right)^2}
            \left[
                \sum_i\sum_j \langle \hat F_i \hat F_j \rangle w_i w_j 
                   - \left(\sum_i \langle\hat F_i\rangle w_i\right)
                    \left(\sum_j \langle\hat F_j\rangle w_j\right)
                \right] \\
        &= \frac{1}{\left(\sum_k w_k\right)^2}
             \left[
                \sum_i\sum_j \langle \hat F_i \hat F_j \rangle w_i w_j 
                - \left(\sum_i \langle\hat F_i\rangle^2 w_i^2
                + \sum_i \sum_{j\ne i} \langle\hat F_i\rangle\langle\hat F_j\rangle w_i w_j\right)
            \right].
    \end{aligned}
    \end{equation}
If the noise in different frequency channels is uncorrelated, then $\langle \hat F_i \hat F_j\rangle = \langle \hat F_i\rangle\langle \hat F_j\rangle$ when $i\ne j$, giving
    \begin{equation}
    \label{eq:weightedVariance}
    \begin{aligned}
    \sigma^2
        &= \frac{1}{\left(\sum_k w_k\right)^2}
            \left[
                \left(
                    \sum_i \langle \hat F_i^2\rangle w_i^2
                    + \sum_i \sum_{i\ne j} \langle\hat F_i\rangle\langle \hat F_j\rangle w_i w_j
                \right)
                - \left(
                    \sum_i \langle\hat F_i\rangle^2 w_i^2
                    + \sum_i \sum_{j\ne i} \langle\hat F_i\rangle\langle\hat F_j\rangle w_i w_j
                \right)
            \right] \\
        &= \frac{
            \sum_i \left(\langle\hat F_i^2\rangle - \langle \hat F_i\rangle^2\right) w_i^2
            }{
            \left(\sum_k w_k\right)^2
            } \\
        &= \frac{\sum_i \sigma^2_i w_i^2}{\left(\sum_k w_k\right)^2},
    \end{aligned}
    \end{equation}
where $\sigma^2_i$ is the variance of $\hat F_i$.
With Eqs. \eqref{eq:weightedMean} and \eqref{eq:weightedVariance}, we can define the expected signal-to-noise ratio (S/N) of a radio signal of flux density $F$:
    \begin{equation}
    \label{eq:weightedSNR}
    \begin{aligned}
    \langle \mathrm{S/N}\rangle
        &= \frac{\langle\hat F\rangle}{\sqrt{\sigma^2}} \\
        &= F \frac{\sum_i w_i}{\sqrt{\sum_j \sigma^2_j w^2_j}}.
    \end{aligned}
    \end{equation}

We can now choose the weights $w_i$ that maximize this signal-to-noise ratio.
Note that Eq. \eqref{eq:weightedSNR} can be rewritten as
    \begin{equation}
    \label{eq:SNRinnerProduct}
    \langle \mathrm{S/N}\rangle
        = \frac{\left( w \,|\, \sigma^{-2}\right)}
            {\sqrt{\left(w\,|\,w\right)}},
    \end{equation}
where we have defined the inner product $\left(a\,|\,b\right) = \sum a_i b_i/\sigma^2_i$ between series $a_i$ and $b_i$.
Just as the dot product between two vectors is maximized when the vectors are parallel, Eq. \eqref{eq:SNRinnerProduct} is maximized when $w_i \propto \sigma^{-2}_i$.
Specifically, if we choose $w_i = \sigma^{-2}_i$, we obtain Eq. \eqref{eq:AppendixFhat} above.

%%%%%%%%%%%%%%%%%%%%%%%%%%%%%%%%%%
\section{Flux and Luminosity Upper Limits}
\label{sec:bayes}
%%%%%%%%%%%%%%%%%%%%%%%%%%%%%%%%%%

Given radio data $d$, our first goal is to compute the posterior $p(F|d,\hat\Omega)$ on the radio flux of GW170104 for every possible sky location.
This posterior is obtained by marginalizing over dispersion measure and initial signal time:
    \begin{equation}
    p(F|d,\hat\Omega)
        = \int d\mathrm{DM}\,\int dt_0\,
            p(F|d,\hat\Omega,\mathrm{DM},t_0) p(\mathrm{DM}) p(t_0),
    \end{equation}
where $p(\mathrm{DM})$ and $p(t_0)$ are our prior probabilities on a signal's dispersion measure and initial time.
Next, using Bayes' theorem, we can relate the posterior $p(F|d,\hat\Omega,\mathrm{DM},t_0)$ to the likelihood $p(d|F,\hat\Omega,\mathrm{DM},t_0)$ of having measured $d$:
    \begin{equation}
    p(F|d,\hat\Omega)
        \propto \int d\mathrm{DM}\,\int dt_0\,
             p(d|F,\hat\Omega,\mathrm{DM},t_0)
             p(F) p(\mathrm{DM}) p(t_0),    
    \end{equation}
where $p(F)$ is our flux density prior.
For every choice of $\hat\Omega$, $\mathrm{DM}$, and $t_0$, we have computed an estimator $\hat F(\hat\Omega,\mathrm{DM},t_0)$ for the corresponding de-dispersed flux, together with an estimate $\sigma^2(\hat\Omega,\mathrm{DM},t_0)$ of its variance.
Assuming Gaussian statistics, the likelihood $p(d|F,\hat\Omega,\mathrm{DM},t_0)$ may be written
    \begin{equation}
    p(d|F,\hat\Omega,\mathrm{DM},t_0)
        \propto \exp \left(
            -\frac{
            \left[\hat F(\hat\Omega,\mathrm{DM},t_0)-F\right]^2}
            {2\sigma^2(\hat\Omega,\mathrm{DM},t_0)}
            \right).
    \end{equation}
Meanwhile, for simplicity we assume flat priors over the ranges $\Delta\mathrm{DM}$ and $\Delta t_0$ considered:
$p(\mathrm{DM}) = 1/\Delta\mathrm{DM}$ and $p(t_0) = 1/\Delta t_0$.
We similarly assume a uniform prior on all positive $F$.
All together,
    \begin{equation}
    \label{eq:fluxPosterior}
    p(F|d,\hat\Omega)
        = \int \frac{d\mathrm{DM}}{\Delta\mathrm{DM}}\,\int \frac{dt_0}{\Delta t_0}\,
             \frac{1}{\mathcal{N}(\hat\Omega,\mathrm{DM},t_0)} \exp \left(
                -\frac{
                    \left[\hat F(\hat\Omega,\mathrm{DM},t_0)-F\right]^2}
                    {2\sigma^2(\hat\Omega,\mathrm{DM},t_0)}
            \right),
    \end{equation}
with normalization factor
    \begin{equation}
    \mathcal{N}(\hat\Omega,\mathrm{DM},t_0)
        = \int_0^\infty dF \exp \left(
            -\frac{
            \left[\hat F(\hat\Omega,\mathrm{DM},t_0)-F\right]^2}
            {2\sigma^2(\hat\Omega,\mathrm{DM},t_0)}
            \right).
    \end{equation}
With the flux posterior $p(F|d,\hat\Omega)$ in hand, the 95\% credible flux upper limit in direction $\hat\Omega$ corresponds to the flux $F_{95}$ satisfying
    \begin{equation}
    \label{eq:fluxLimit}
    0.95 = \int_0^{F_{95}} dF p(F|d,\hat\Omega).
    \end{equation}
These upper limits are shown in Fig. \ref{fig:limits} above.
    
Using the Advanced LIGO posterior on GW170104's location, we can additionally compute a posterior $p(L|d)$ on the equivalent isotropic luminosity of GW170104, marginalized over all possible progenitor sky locations $\hat\Omega$ and distances $D$:
    \begin{equation}
    p(L|d) = \int dD \int d\hat\Omega\, p(L|d,D,\hat\Omega) p(D,\hat\Omega).
    \label{eq:post1}
    \end{equation}
Here, $p(D,\hat\Omega)$ is the probability distribution on the progenitor location of GW170104; we take this to be the localization provided by Advanced LIGO.
As in Eq. \eqref{eq:fluxLimit} above, the 95\% credible upper limit is given by the luminosity $L_{95}$ satisfying $0.95 = \int_0^{L_{95}} p(L|d) dL$, or
    \begin{equation}
    0.95 = \int dD \int d\hat\Omega\int_0^{L_{95}} dL\, p(L|d,D,\hat\Omega) p(D,\hat\Omega).
    \label{eq:post2}
    \end{equation}
Note that, as currently written, this equation requires posterior probabilities $p(L|d,D,\hat\Omega)$ on luminosity as a function of direction and distance.
We can recast Eq. \eqref{eq:post2} in terms of our known flux posteriors $p(F|d,\hat\Omega)$ (Eq. \eqref{eq:fluxPosterior}) by substituting $p(L|d) = p(F|d)\,dF/dL = p(F|d)/4\pi D^2$ and $dL = 4\pi D^2 dF$, giving
    \begin{equation}
    0.95 = \int dD \int d\hat\Omega\int_0^{F(D,L_{95})} dF\,p(F|d,\hat\Omega) p(D,\hat\Omega).
    \label{eq:post3}
    \end{equation}
In practice, Eq. \eqref{eq:post3} is somewhat easier to evaluate when rearranged as 
    \begin{equation}
    0.95 = \int d\hat\Omega\,p(\hat\Omega)
        \int dD\,p(D|\hat\Omega)
        \int_0^{F(D,L_{95})} dF\,p(F|d,\hat\Omega).
    \label{eq:ul1}
    \end{equation}

\end{document}